# Mapping the full lattice strain tensor of a single dislocation by High Angular Resolution Transmission Kikuchi Diffraction (HR-TKD)

Hongbing Yu[1], Junliang Liu[2], Phani Karamched[2], Angus J. Wilkinson[2], Felix Hofmann[*1]

(1) Department of Engineering Science, University of Oxford, Parks Road, Oxford, OX1 3PJ, UK

(2) Department of Materials, University of Oxford, Parks Road, Oxford, OX1 3PH, UK

**Abstract**

The full lattice strain tensor and lattice rotations induced by a dislocation in pure tungsten were mapped using high resolution transmission Kikuchi diffraction (HR-TKD) in a SEM. The HR-TKD measurement agrees very well with a forward calculation using an elastically isotropic model of the dislocation and its Burgers vector. Our results demonstrate that the spatial and angular resolution of HR-TKD in SEM is sufficiently high to resolve the details of lattice distortions near individual dislocations. This capability opens a number of new interesting opportunities, for example determining the Burgers vector of an unknown dislocation in a fast and straightforward way.

Electron backscatter diffraction (EBSD) in a scanning electron microscope (SEM) is widely applied in material characterisation at the mesoscale. By rastering a focused electron beam across a grid of points on the sample surface and analyzing EBSD patterns, the crystallographic orientation of each point is obtained [1,2]. Based on the point-by-point orientation, information such as grain structure [3], phase identification [3], intragranular misorientations [1,4,5], micro-texture [4,6], grain boundaries [4–7] and orientation relationships between phases [1] can be retrieved. The angular resolution of EBSD is ~1° [8]. The cross-correlation based EBSD analysis approach introduced by Wilkinson (HR-EBSD) improves the angular resolution to 0.005° by measuring small shifts of features in the EBSD patterns compared to a reference EBSD pattern [8–11]. These small shifts can be interpreted in terms of lattice rotations and lattice distortions [1,8–11].

---

[*] Corresponding author: felix.hofmann@eng.ox.ac.uk

HR-EBSD has been widely adopted to characterise geometry necessary dislocation (GND) density and residual lattice strains in crystalline materials [1,12–17]. A comparative study showed good agreement between HR-EBSD and X-ray measurements of GND density [13]. Importantly, HR-EBSD can access the spatial distribution of GND density and lattice strain at the nano-scale near interesting features, such as grain boundaries [18–20], indents [21,22], second phases [20,23] or slip bands [24]. The spatial resolution of HR-EBSD is governed by the electron interaction volume with estimates of the probed volume ranging from several tens to hundred nanometers in bulk material [25]. This, and experimental issues with drift, have prevented the study of the strain fields associated with individual dislocations using HR-EBSD, though statistical analysis by Wilkinson et al [26] indicates that sufficient spatial resolution should be available to probe the lattice strains near dislocations.

Dislocations are one of the most important lattice defects in crystalline materials. Thus far detailed characterisation of dislocations has mostly relied on TEM for determination of dislocation type, Burgers vector (**b**) and associated strain fields. The two most common methods for measuring lattice strain at the atomic scale are geometric phase algorithms (GPA) [27,28] and nano-beam diffraction in TEM [29–31]. Both offer a strain sensitivity of $\sim 10^{-3}$ and a spatial resolution of 2 to 3 nm [32]. However, only the 2D in-plane strain tensor can be measured from these techniques, and the measurement must be performed on a certain zone axis, placing stringent requirements on sample preparation.

By using transmission Kikuchi diffraction [25] (TKD) in a SEM, i.e. detecting the Kikuchi pattern from the bottom surface of a thin foil, the absolute spatial resolution can be improved to ~10 nm [25], while the effective spatial resolution falls to 2-4 nm [33] . Here we show that by combining TKD with HR-EBSD approaches in the SEM, it becomes possible to measure the full deviatoric strain tensor associated with an individual dislocation. These measurements are compared to predicted strain fields, calculated using an isotropic elasticity model. Our results show that HR-TKD provides a convenient and reliable way of probing nano-scale strain fields, with sufficient sensitivity to study strains associated with specific dislocations.

Ultra-high purity tungsten foil (99.99% in purity and 120 μm thick) was punched into 3 mm diameter discs. The samples were thinned to electron transparency by twin-jet electropolishing (0.5 wt% NaOH aqueous solution, 0 °C, 14 V). **g·b** analysis was performed on a JEOL 2100 TEM. TEM bright field images under 8 independent **g** vectors from 4 zone axes were acquired. A Zeiss Merlin SEM with a Bruker eFlash detector was then used to carry out HR-TKD measurements (20 kV, 3 nA). The TKD setup is shown in Fig. 1 (a). The sample was tilted -45° to the electron beam, a TKD pattern size of 800 ×600 selected and a scanning step size of 4 nm used. The cross-correlation analysis of the Kikuchi patterns was done by the XEBSD matlab code described by Britton & Wilkinson [34,35].

The anticipated spatial variation of the deviatoric lattice strain tensor and lattice rotations in the vicinity of dislocation were calculated using isotropic elasticity. This is reasonable since tungsten is almost perfect elastically isotropic [36]. Since a thin foil was used for the measurement, the surface relaxation was taken into account. For simplicity, we assume that the dislocation is straight with line direction normal to the foil surface, i.e. along the -z direction (see supplementary Fig. S1). For a dislocation with arbitrary Burgers vector **b** that meets the free surface at (0, 0, 0), the displacement field at (x, y, z) can be found by decomposing the Burgers vector into components along x, y, z directions:

$$\mathbf{b} = \sum_{i=1}^{3} b_i \mathbf{n}_i, \tag{1}$$

where $\mathbf{n}_i$ are unit vectors along x, y, and z directions, $b_i$ are the corresponding coefficients.

The displacement field caused by the dislocation with Burgers vector **b** is then the linear superposition of the displacements caused by a screw dislocation with Burgers vector $b_3 \mathbf{n}_3$ and two edge dislocations with Burgers vectors $b_1 \mathbf{n}_1$ and $b_2 \mathbf{n}_2$. The displacement field can then be expressed as,

$$u_{x_i} = \sum_{j=1}^{3} u_{x_i}^{j} \quad (j = 1, 2, 3), \tag{2}$$

where $u_{x_i}$ is total displacement along the $x_i$ direction ($x_i$ refers to *x, y* or *z*). $u_{x_i}^{j}$ is the displacement along $x_i$ induced by dislocation with Burgers vector $b_j \mathbf{n}_j$.

The displacement field of an edge or screw dislocation meeting a free surface can be obtained by superposing the displacement of the dislocation in an infinite body and the displacement induced by image forces due to the traction free surface condition,

$$u_{x_i}^j = u_{x_i}^{j\_inf} + u_{x_i}^{j\_image} \quad (i,j = 1,2,3). \tag{3}$$

Here we use the solution for $u_{x_i}^{j\_inf}$ and $u_{x_i}^{j\_image}$ given by Anderson et al [37] and Yoffe [38] (also provided in the supplementary material). The components of the 3D strain tensor of the dislocation at any position (x, y, z) are obtained by differentiation:

$$\varepsilon_{ij} = \frac{1}{2}\left(\frac{\partial u_{x_i}}{\partial x_j} + \frac{\partial u_{x_j}}{\partial x_i}\right) \quad (i,j = 1,2,3). \tag{4}$$

The lattice rotation can be obtained as

$$\omega_{ij} = \frac{1}{2}\left(\frac{\partial u_{x_i}}{\partial x_j} - \frac{\partial u_{x_j}}{\partial x_i}\right) \quad (i,j = 1,2,3). \tag{5}$$

As the TKD patterns are dominated by a tens of nanometers thick surface layer, the reported elastic strain and lattice rotations are an average over a depth of 50 nm.

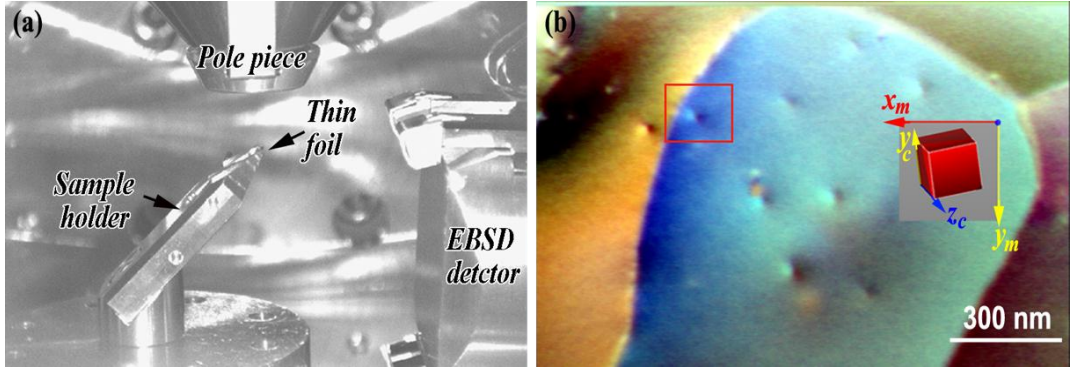

**Fig. 1** (a) The setup for HR-TKD measurement. (b) transmitted primary electron image showing the contrast from dislocations. The inset cube shows the orientation of the grain. $x_m$-$y_m$ is the coordinate frame of the electron image and the EBSD map. $y_c$-$z_c$ is the crystal coordinate frame.

In SEM, dislocations can be imaged with high-energy primary electrons (PE) without needing to set up specific diffraction conditions. Fig. 1 (b) shows a PE image of the grain under investigation. Several dislocations with line direction near normal to the surface can be seen with black-white contrast. The crystal orientation in the grain, determined by TKD, is (227.8°, 23.2°, 322.3°) in Euler angles and the corresponding coordinate frames for the image and the crystal are shown inset in Fig. 1 (b). A dislocation close to a grain boundary (2~3° misorientation) was selected for TEM **g· *b*** analysis and HR-TKD (red rectangle in Fig. 1 (b)) as the grain boundary provides a convenient reference for judging drift during the HR-TKD measurement.

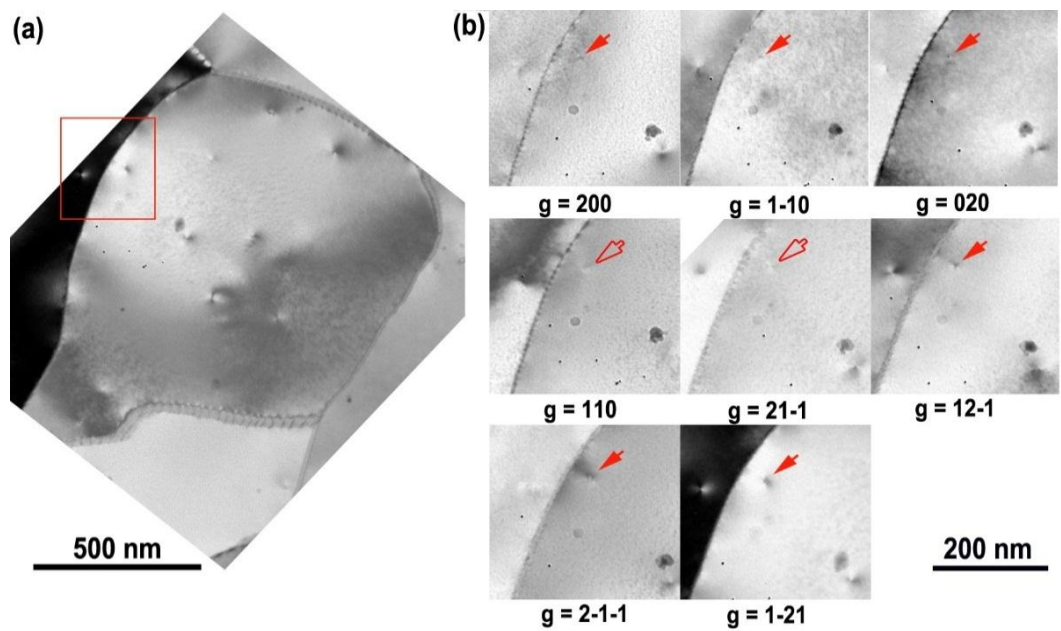

**Fig. 2** (a) TEM bright field image of the grain shown in Fig.1 (b). (b) **g· b** analysis. Red arrows point to the dislocation under study. Full and hollow arrows indicate visibility and invisibility of the dislocation respectively.

Fig. 2 (a) shows the same grain imaged by TEM bright field. The dislocations imaged by the two different techniques agree very well. **g· b** analysis for the dislocation of interest is shown in Fig. 2 (b). The **g** vectors were determined consistent with the crystal orientation found by EBSD. The dislocation shows contrast except under **g = 110** and **g = 21$\bar{1}$** conditions. The

Burgers vector of the dislocation can thus be determined as $\pm[1\bar{1}1]/2$ according to the **g· b** table (Table 1).

**Table 1** The **g·b** table for visibility (v) and invisibility (i) of dislocations in bcc crystal.

| b \ g | 200 | $1\bar{1}0$ | 020 | 110 | $(21\bar{1})$ | $(12\bar{1})$ | $(2\bar{1}\,\bar{1})$ | $(1\bar{2}1)$ |
|---|---|---|---|---|---|---|---|---|
| $\pm[100]$ | v | v | i | v | v | v | v | v |
| $\pm[010]$ | i | v | v | v | v | v | v | v |
| $\pm[001]$ | i | i | i | i | v | v | v | v |
| $\pm[111]$ | v | i | v | v | v | v | i | i |
| $\pm[11\bar{1}]$ | v | i | v | v | v | v | v | v |
| $\pm[1\bar{1}1]$ | v | v | v | i | i | v | v | v |
| $\pm[\bar{1}11]$ | v | v | v | i | v | i | v | v |

Strain maps for the 6 components of the 3D strain tensor (upper triangle and diagonal in the matrix map) and 3 lattice rotations (lower triangle in the matrix map) near the dislocation, measured by HR-TKD, are shown in Fig. 3(a). The strains and rotations are plotted in the microscope coordinate frame shown in the upper left corner, which is the same as used in Fig. 1 and 2. The deviatoric, rather than full, lattice strain tensor is measured, as HR-TKD is not sensitive to lattice dilation [8]. However, the volumetric strain can be calculated by assuming stress along the out of plane direction to be zero (see supplementary Fig. S2). From the normal strain components, the contrast formed by compressive and tensile strains on either side of the dislocation is clearly visible. The core of the dislocation can be determined as the middle point between peaks of compressive strain and tensile strain in the $\varepsilon_{22}$ strain map, which is the clearest of the measured lattice strain components. The core is marked by two intersecting arrows superposed on the $\varepsilon_{12}$ strain map. The strain maps for shear components of $\varepsilon_{12}$ and $\varepsilon_{13}$ are a little noisy, while the map for $\varepsilon_{23}$ shows obvious contrast between positive and negative shear strains. Lattice rotations around axes $z_m$ ($\omega_{12}$) and $x_m$ ($\omega_{23}$) display clear negative and positive contrast between the two sides of the dislocation, while the lattice rotations around axis $y_m$ ($\omega_{13}$) are small.

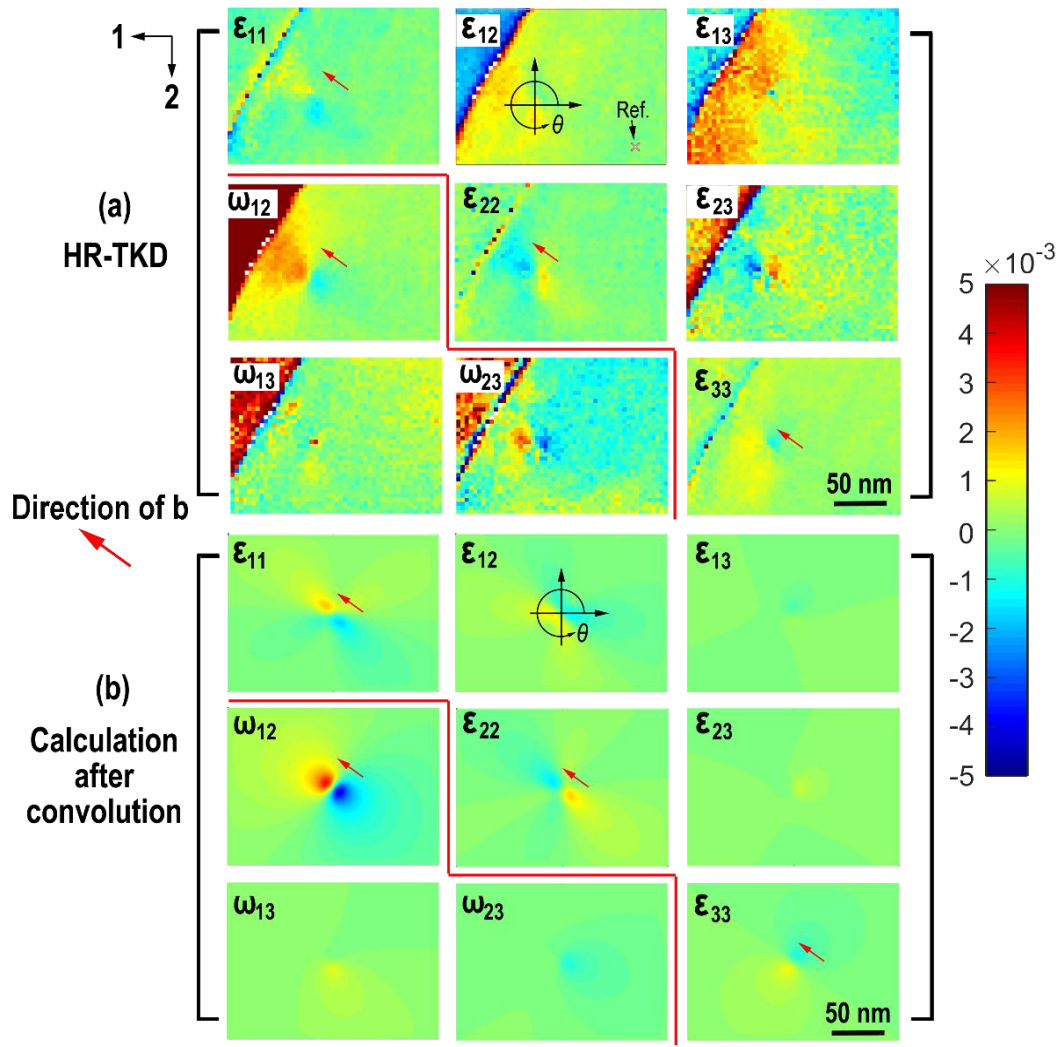

**Fig. 3** (a) Maps for 3D lattice strain tensor and lattice rotations around the dislocation measured by HR-TKD. (b) The forward calculation of the distribution of strain tensor and lattice rotation around a $[\bar{1}1\bar{1}]/2$ dislocation using an isotropic elastic forward model convolved with a 2D Gaussian probe function. Red arrows shows the direction of Burgers vector projection. The cross shown in the $\varepsilon_{12}$ map is the place where reference pattern was taken.

It is interesting to note that the separation between the tensile and compressive strain peak in the $\varepsilon_{22}$ strain map is ~25nm. The isotropic elasticity model, on the other hand, predicts no separation between these maxima (see Supplementary Fig. S3). The elasticity model has a singularity at the dislocation core, however, at >10 b (~2.7 nm) away from the core [39], elasticity is valid. This suggests that the large separation between extreme $\varepsilon_{22}$ values we observe is due to the finite size of the electron interaction volume. To enable a better comparison between the measured and predicted strain fields, the forward calculated lattice strain and lattice rotation maps were convolved with a 2D Gaussian function (σ = 5 nm). This estimate of the probe resolution is consistent with previous reports of TKD spatial resolution [25]. Fig. 3(b) shows the forward calculation of the variation of deviatoric strain tensor and lattice rotations near the dislocation after convolution with the probe function. Importantly only the strain fields for **b** = [$\bar{1}1\bar{1}$]/2 correctly capture the spatial variation of strains and rotations. The alternative **b** = [$1\bar{1}1$]/2, which is also consistent with the **g.b** analysis, would yield the opposite, incorrect, sign of all strain and rotation components.

Good agreement between the measurement and the calculation can be found in all the components except $\varepsilon_{13}$, $\varepsilon_{23}$, and $\omega_{23}$. Particularly we note that key features, such as the line separating positive and negative lobes, as well as the direction pointing from the negative to the positive lobes match remarkably well (see the maps for $\varepsilon_{11}$, $\varepsilon_{22}$, $\varepsilon_{33}$, $\varepsilon_{12}$, and $\omega_{12}$ ). It is noted that in the $\varepsilon_{22}$ map the direction from the positive to negative lobes is exactly parallel to the projection of **b**, while the negative to positive direction in $\varepsilon_{11}$ map is a little bit off from the projection of **b**. This matches perfectly with the calculation.

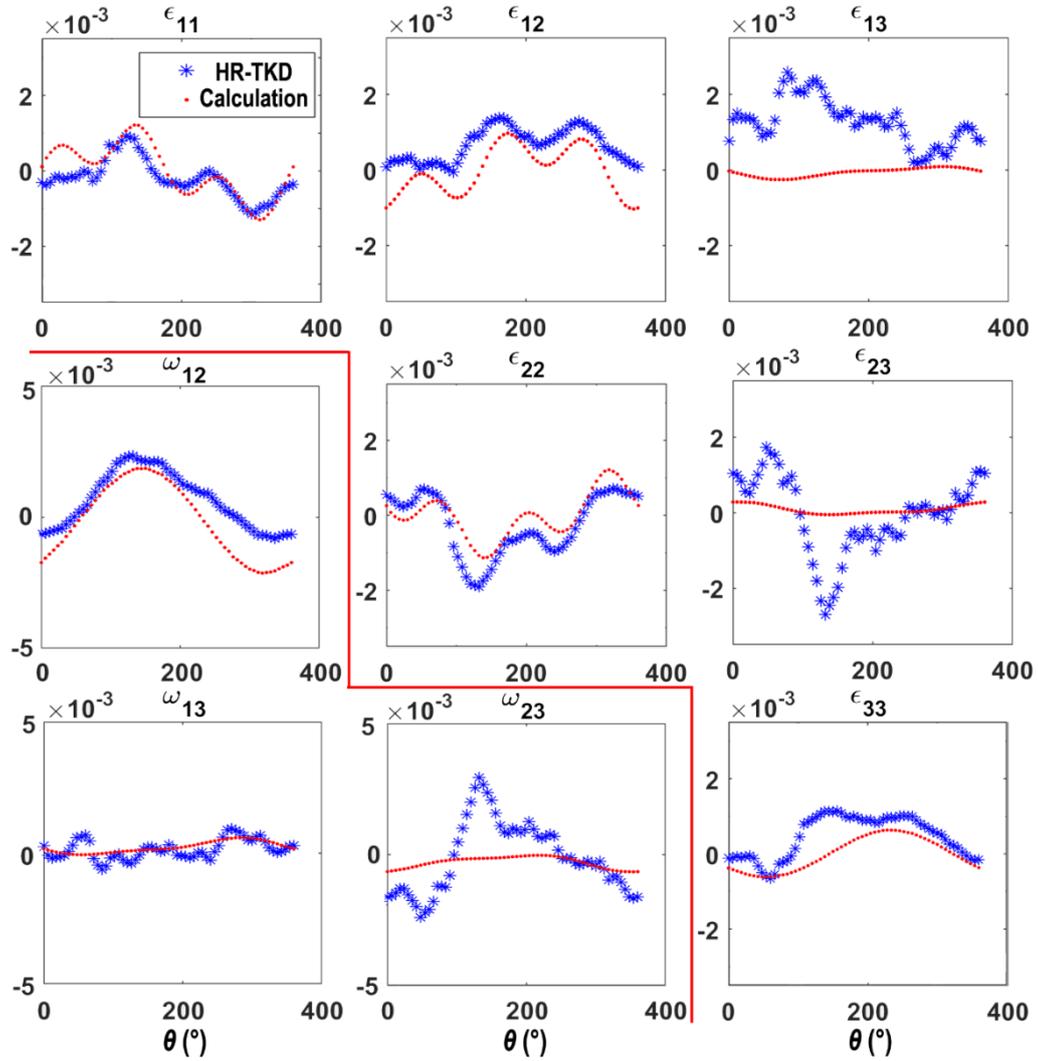

**Fig. 4** Angular variation of the lattice strain field and lattice rotation field around a circular path at a 20 nm radius from the dislocation core. The path considered is shown in Fig. 3 superimposed on the $\varepsilon_{12}$ strain component.

The agreement can be further quantified by considering the angular variation of the lattice strain and lattice rotation fields along a circular path around the dislocation 20 nm from the core (see path drawn on $\varepsilon_{12}$ maps in Fig. 3). The resulting profiles for all strain components are shown in Fig. 4. For the direct strain components ($\varepsilon_{11}$, $\varepsilon_{22}$, $\varepsilon_{33}$), the $\varepsilon_{12}$ shear component and $\omega_{12}$ and $\omega_{13}$ lattice rotations the measurement agrees rather closely with the expected angular variation. In particular, it is very interesting that the location of key features, such as peaks and troughs are correctly captured. Several of the profiles appear to have a small vertical offset. This may be explained as the lattice rotations and strains in HR-TKD are computed with respect to a reference pattern assumed to come from a nominally strain free region of the sample.

The large difference found in $\varepsilon_{13}$ is presumably caused by the grain boundary, as it can be seen that a large positive strain is seen along the grain boundary. Surprisingly the experimental measurements of $\varepsilon_{23}$ and $\omega_{23}$ show obvious positive and negative contrast around the dislocation, while only small variations of these components are expected from calculations. Qualitatively, the directions from negative to positive lobes match well in measurements and calculations from both $\varepsilon_{23}$ and $\omega_{23}$ (better visible in Supplementary Fig. S3). Considering equations (4 & 5), summing of $\varepsilon_{23}$ and $\omega_{23}$ provides a map of $\frac{\partial u_y}{\partial z}$, which is small, as expected from calculations. This suggests that the large signal in the experimental maps of $\varepsilon_{23}$ and $\omega_{23}$ is due to $\frac{\partial u_z}{\partial y}$ (see supplementary Fig. S4).

The good agreement of the strain tensor and lattice rotation between measurement and calculation shows that the spatial resolution of HR-TKD, as well as the angular resolution is sufficient to study in detail the lattice distortions caused by individual dislocations. Importantly this characterisation can be done in the SEM, allowing the determination of dislocation Burgers vector, which was previously only possible by TEM. In TEM Burgers vector analysis commonly relies on **g·b** analysis and black-white contrast. However, **g·b** analysis becomes ambiguous when applied to small irradiation induced dislocation loops and dislocations normal to the thin foil. Here, because of surface relaxation, dislocations will still show contrast even when **g·b** = 0 [40–42]. Black-white contrast relies on computation of the dynamical diffraction images of dislocations recorded at different **g** vectors. This comparison becomes somewhat involved as a number of other factors will also modify dislocation contrast, increasing complexity of this type of analysis [42]. Instead, direct comparison of full strain tensor and lattice rotations (which actually cause the intensity contrast in TEM) from just one HR-TKD measurement with predicted strain maps provides a straightforward approach to unambiguously determining the Burgers vector direction, sign and magnitude of unknown dislocations. To illustrate this point, the calculated strain tensors for all other possible Burgers vectors in tungsten are shown in supplementary Figs. S5-10. No agreement can found between those maps and the measured maps. Only the correct Burgers vector, $[\bar{1}1\bar{1}]/2$, provides a good match to the measured strain profiles.

In summary, we have demonstrated that, using tungsten as a case study, the full deviatoric lattice strain tensor and rotation field due to an individual dislocation can be quantitatively mapped using HR-TKD. The experimentally measured lattice distortions are in remarkably good agreement with those expected from a forward calculation using an isotropic elasticity model of the dislocation. Our results suggest that the combination of strain field simulation and HR-TKD may offer a straightforward approach to determining Burgers vector magnitude, direction and sign. In principle, this is similar to black-white contrast simulations used in TEM, but rather than interpreting the intensity contrast caused by the strain fields, the strain itself is directly used for the analysis.

**Acknowledgement**

We thank Edmund Tarleton for helpful discussions and Xiaoou Yi for providing the raw material. This work was funded by The Leverhulme Trust under grant RPG-2016-190, and EPSRC under grant EP/K034332/1. Electron microscopy was performed at the David Cockayne Centre for Electron Microscopy at the Department of Materials, University of Oxford.

# Supplementary materials for 'Mapping the full lattice strain tensor of a single dislocation by High Angular Resolution Transmission Kikuchi Diffraction (HR-TKD)'

## 1. Strain Field Calculation for dislocations at a free surface.

The coordinate convention used for simulations of the dislocation displacement fields is shown in supplementary Fig. S1. The displacement field of a screw dislocation ($\mathbf{b} = b_3 \mathbf{n_3}$, $\mathbf{n_3}$ is a unit vector along $z$ direction) in an infinite medium is

$$u_x^{3\_inf} = u_y^{3\_inf} = 0,$$

$$u_z^{3\_inf} = b_3 \frac{tan^{-1}(y/x)}{2\pi}.$$

The displacement field caused by the relaxation of a screw dislocation ($\mathbf{b} = b_3 \mathbf{n_3}$) normal to the free surface, due to the traction free boundary condition, has been provided by Yoffe [1],

$$u_x^{3\_image} = \frac{b_3 y}{2\pi(R-z)},$$

$$u_y^{3\_image} = \frac{b_3 x}{2\pi(R-z)},$$

$$u_z^{3\_image} = 0,$$

where $R = \sqrt{x^2 + y^2 + z^2}$, and $b_3$ is the magnitude of the Burgers vector $\mathbf{b}$.

The total displace field of the screw dislocation at (x, y, z) is

$$u_{x_i}^3 = u_{x_i}^{3\_inf} + u_{x_i}^{3\_image} \quad (i = 1, 2, 3),$$

here $x_1$, $x_2$, $x_3$ represent x, y, z coordinates system.

The displacement field of an edge dislocation ($\mathbf{b} = b_1 \mathbf{n_1}$, $\mathbf{n_1}$ is unit vector along $x$ direction) in an infinite medium is

$$u_x^{1\_inf} = \frac{b_1}{2\pi}\left[(tan^{-1}\left(\frac{y}{x}\right) + \frac{xy}{2(1-v)(x^2+y^2)}\right],$$

$$u_y^{1\_inf} = \frac{-b_1}{2\pi}\left[\frac{1-2v}{4(1-v)}ln(x^2+y^2) + \frac{x^2-y^2}{4(1-v)(x^2+y^2)}\right],$$

$$u_z^{1\_inf} = 0,$$

where $v$ is poison ratio.

The displacement field caused by surface relaxation of an edge dislocation ($\mathbf{b} = b_1\mathbf{n_1}$) normal to the free surface is

$$u_x^{1\_image} = \frac{vb_1}{4\pi(1-v)}\left[\frac{2xyz}{R(R-Z)^2} + \frac{(1-2v)xy}{(R-Z)^2}\right],$$

$$u_y^{1\_image} = \frac{vb_1}{4\pi(1-v)}\left[(1-2v)ln(R-z) - \frac{(3-2v)z}{R-z} + \frac{(3-2v)y^2}{(R-z)^2} - \frac{2y^2}{R(R-z)}\right],$$

$$u_z^{1\_image} = \frac{vb_1 y}{2\pi(1-v)}\left(\frac{1}{R} + \frac{1-2v}{R-Z}\right).$$

The total displacement field of the edge dislocation at the surface is

$$u_{x_i}^1 = u_{x_i}^{1\_inf} + u_{x_i}^{1\_image} \quad (i = 1,2,3).$$

In a similar way, we can obtain the total displacement field, $\mathbf{u}_{x_i}^2$, of an edge dislocation with $\mathbf{b} = b_2\mathbf{n_2}$ ($\mathbf{n_2}$ is unit vector along $y$ direction) and normal to the surface.

## 2. Supplementary figures

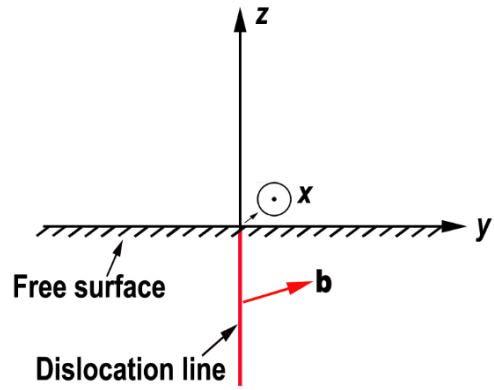

**Fig. S1** The coordinate system setup for electron diffraction measurements and simulations of the elastic strain fields associated with the dislocation.

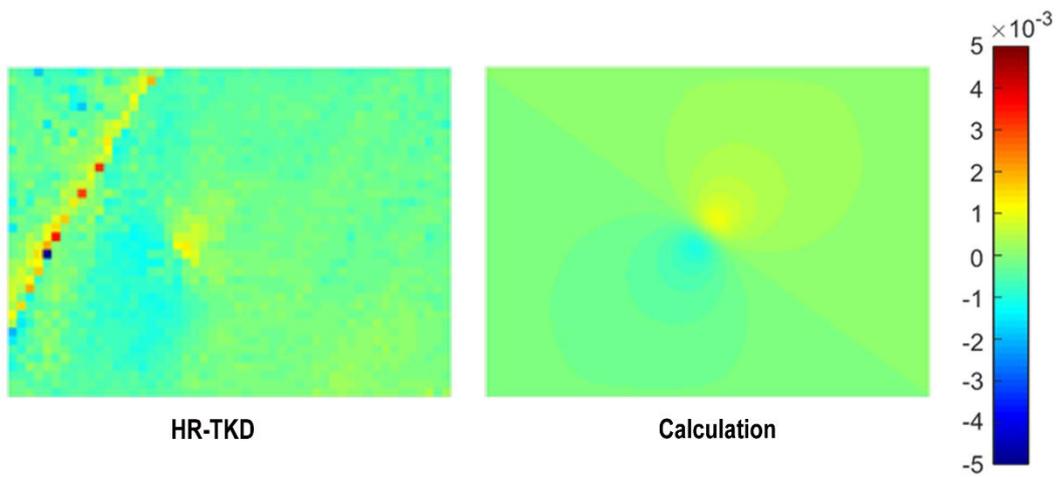

**Fig. S2** The volumetric strain indirectly calculated from HR-TKD measurement (left) and the elasticity calculation (right).

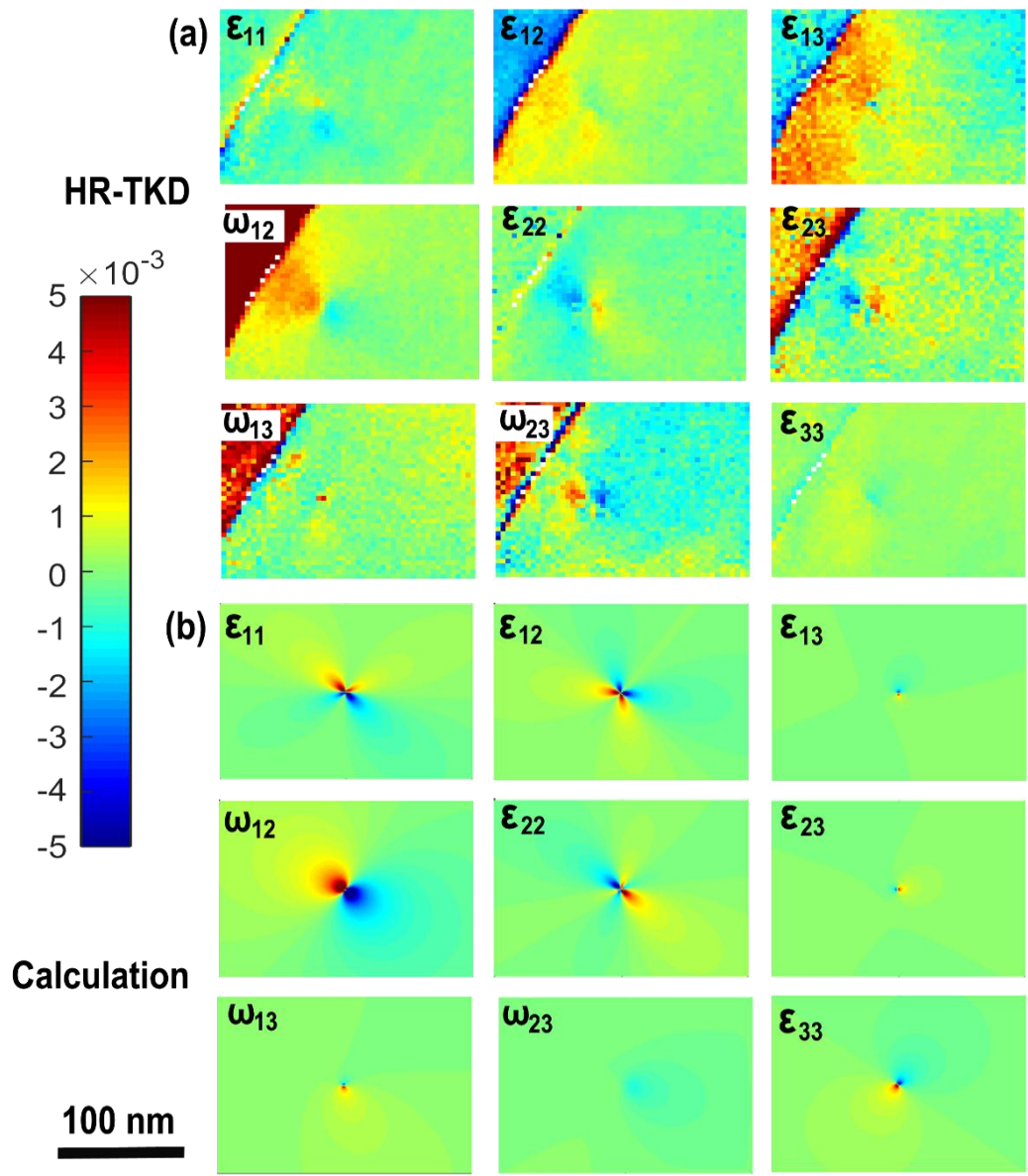

**Fig. S3** The comparison between the HR-TKD and elasticity model before convolution.

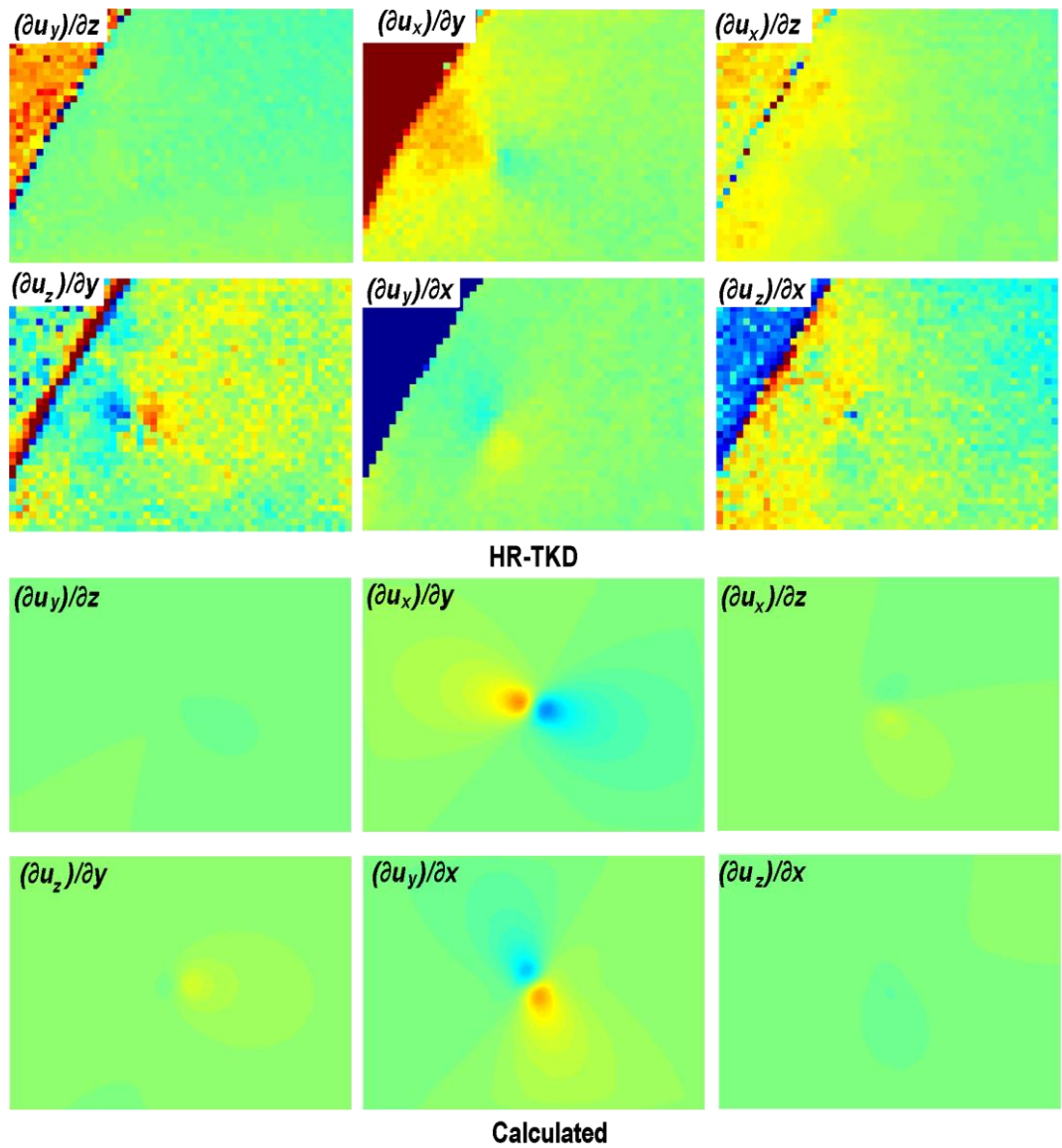

**Fig. S4** Comparison between the HR-TKD measurement and prediction for individual spatial derivatives (b = $[\bar{1}1\bar{1}]/2$). The colour code is the same as Fig. S3.

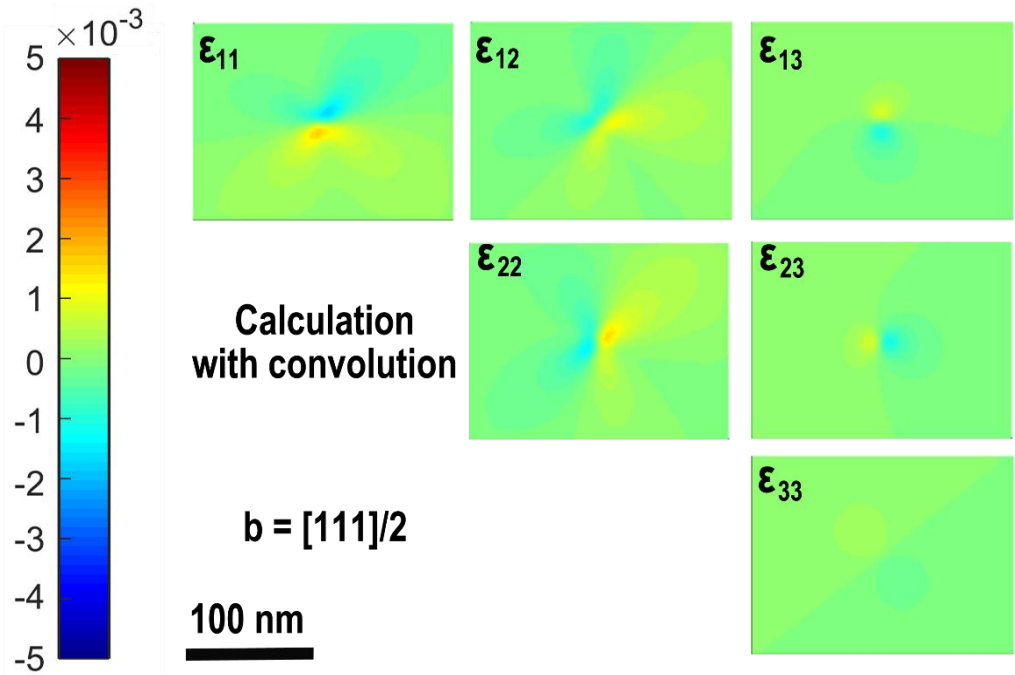

**Fig. S5** The strain tensor expected for the dislocation with b = [111]/2.

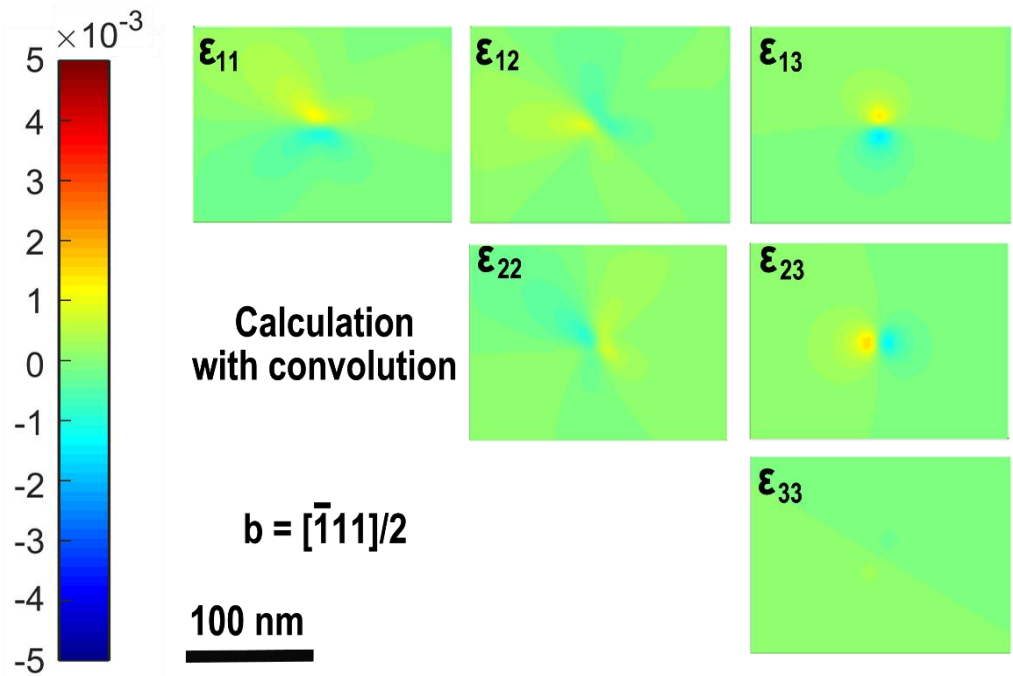

**Fig. S6** The strain tensor expected for the dislocation with b = [$\bar{1}$11]/2.

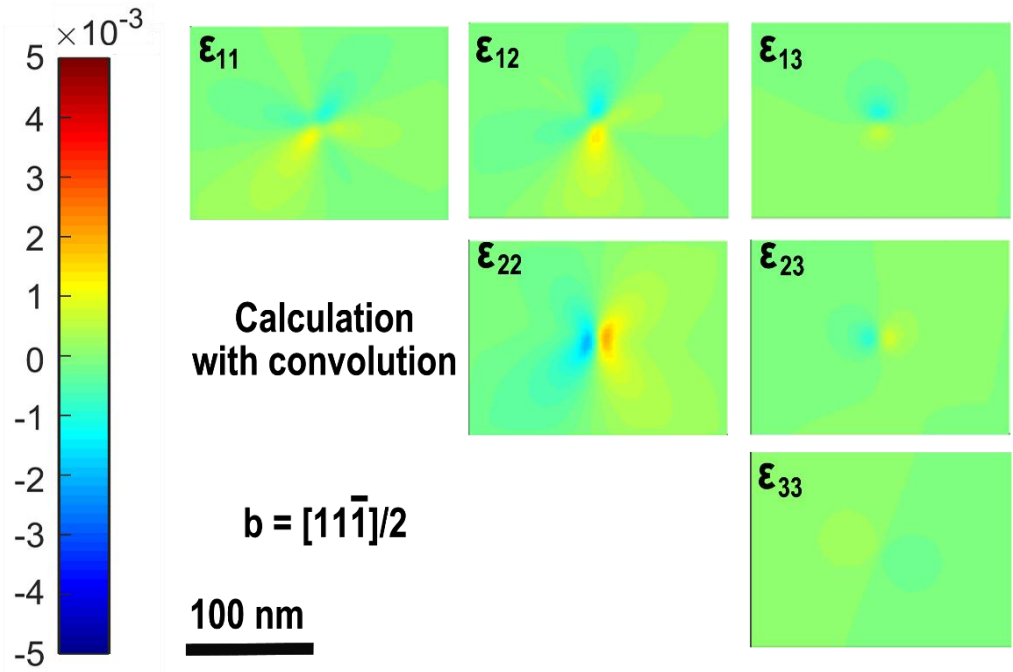

**Fig. S7** The strain tensor expected for the dislocation with b = $[11\bar{1}]/2$.

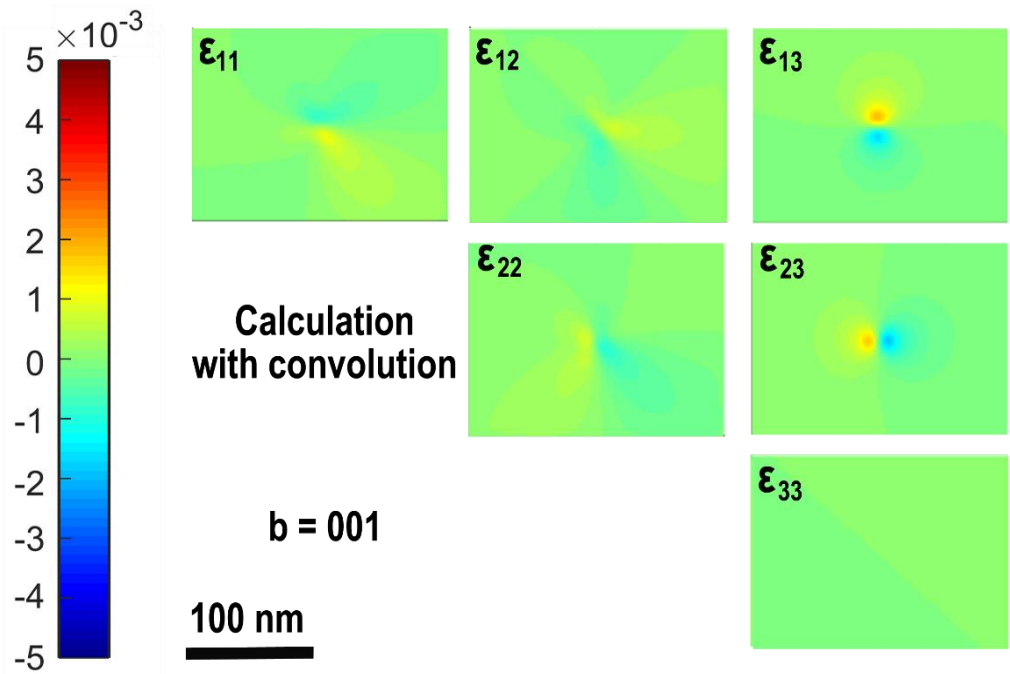

**Fig. S8** The strain tensor expected for the dislocation with b = [001].

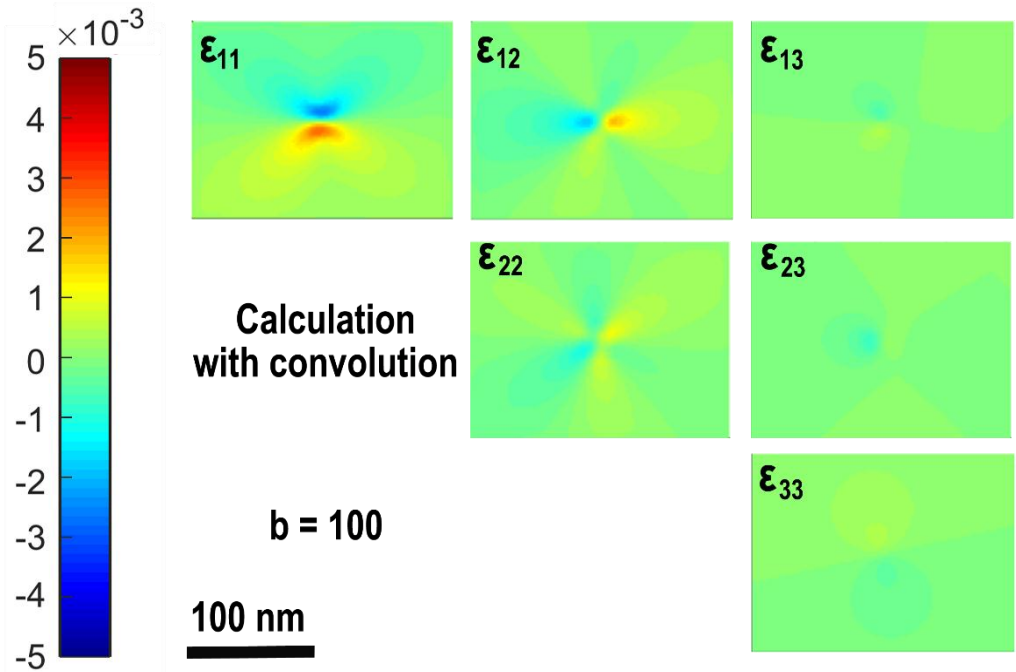

**Fig. S9** The strain tensor expected for the dislocation with b = [100].

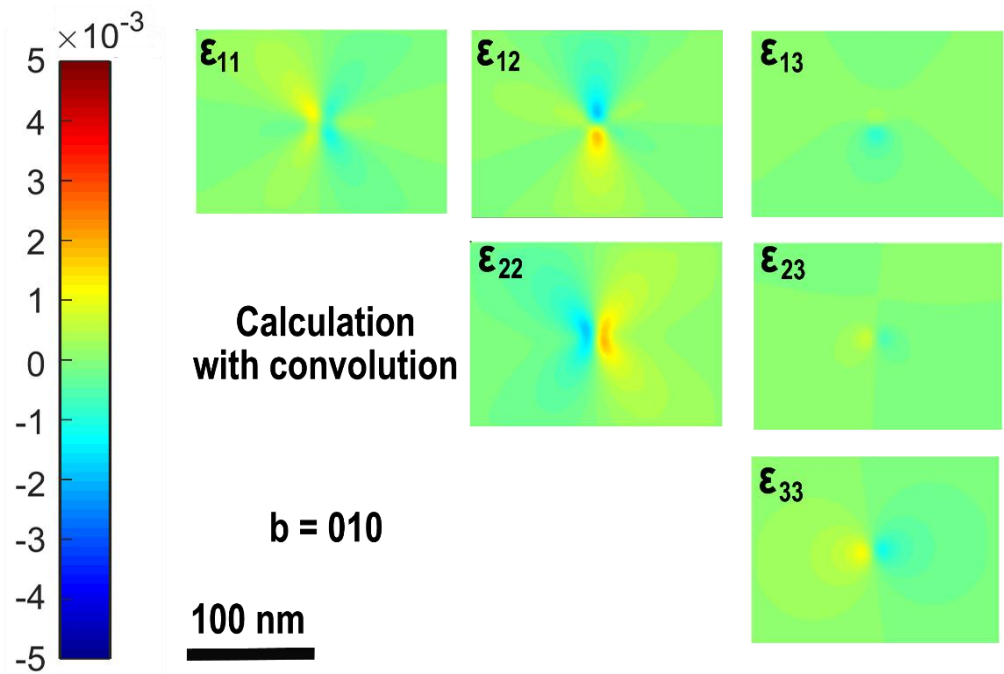

**Fig. S10** The strain tensor expected for the dislocation with b = [010].